\documentclass[twocolumn,preprintnumbers,amsmath,amssymb]{revtex4-2}
\usepackage{graphicx}
\usepackage{color}
\usepackage{caption}
\usepackage{subfigure}
\begin{document}

\title{The nature of water interactions and the molecular signatures of hydrophilicity}

\author{Nicol\'as A. Loubet$^1$}
\author{Alejandro R. Verde$^1$}
\author{Gustavo A. Appignanesi$^{1 *}$}

\affiliation{
$^1$ INQUISUR, Departamento de Qu\'{i}mica, Universidad Nacional del Sur (UNS)-CONICET, Avenida Alem 1253, 8000 Bah\'{i}a Blanca, Argentina\\
* Corresponding author: appignan@criba.edu.ar\\
}

\date{\today}

\begin{abstract}

The peculiar structuring of liquid water stems from a fine-tuned molecular principle embodying the two different interaction demands of the water molecule: The formation of hydrogen bonds or the compensation for coordination defects. Here we shall show that the same underlying molecular mechanism is also at play in order to establish favorable interactions with other systems. In this regard, the emergence of two limiting behaviors for hydrophilicity will help both to unravel its molecular underpinnings and to establish absolute values for such property.

\end{abstract}


\maketitle

Contrary to what one would expect from the austere beauty of its chemical formula, liquid water exhibits quite a complex behavior marked by a plethora of structural, thermodynamic and dynamic anomalies\cite{ball,chaplin,water_chemrev,water1_epje,water2_epje,tanaka,criticalpoint,Kim,francesco-pablo}. Besides its intrinsic relevance, such behavior becomes paramount in main contexts where water plays an active determining role\cite{debenedetti,graphene-hydrophilic,Fluid-Phase-Equilibria,PLOS-con-AF,AFlibro,annurev_Patel,annurev_chembioeng,SAMS_exp,Garde_PNAS,membrane_Martelli,Marcia,Bordin-GAA,Granick} to the point, for example, of being irreplaceable in biology. However, its comprehension is still far from being complete. In particular, understanding and predicting the way in which water interacts with surfaces of different nature is a problem of fundamental interest which, in turn, might impact strongly on central fields like protein folding and protein binding, self-assembly processes in biology and materials science, hydration at interfacial or nanoconfined environments, adsorption, catalysis and water conduction, among many others\cite{debenedetti,graphene-hydrophilic,Fluid-Phase-Equilibria,PLOS-con-AF,AFlibro,annurev_Patel,annurev_chembioeng,SAMS_exp,Garde_PNAS,membrane_Martelli,Marcia,Bordin-GAA,Granick}. This knowledge might also help advance rational design efforts for relevant applications in such realms. The affinity that a given system exhibits for water is conventionally termed  ``hydrophilicity'' and is opposite to  ``hydrophobicity'', which implies a lack of water affinity\cite{debenedetti,annurev_Patel,annurev_chembioeng,Granick,SAMS_exp,Garde_PNAS}. Hydrophilic behavior stems from the ability of the system to establish favorable interactions with water, whose hydrogen-bond (HB) network could be defied by the interacting process. The usual way to define the hydrophilicity of a given system (generally a solid) is to measure the contact angle formed between the tangent line of a water droplet with the surface line (in this case we will work with the external angle): if the contact angle is lower than $\theta=90^{\circ}$, the system is considered as hydrophilic; else, the behavior would be hydrophobic. However, in the absence of a molecular descriptor acting as a marker of the transition between hydrophilic and hydrophobic behavior, such value of $\theta$ remains as a merely arbitrary definition (nothing noticeable has been so far detected to happen as this limit is crossed, as recognized in a popular commentary\cite{Granick}). Given this lack of a profound understanding of the molecular nature of hydrophilicity, it is not surprising that hydrophobicity scales that already exist in different contexts have been built mainly on a relative basis rather than on absolute values. 

\begin{figure}[h!]
\resizebox{0.5\textwidth}{!}{%
\includegraphics{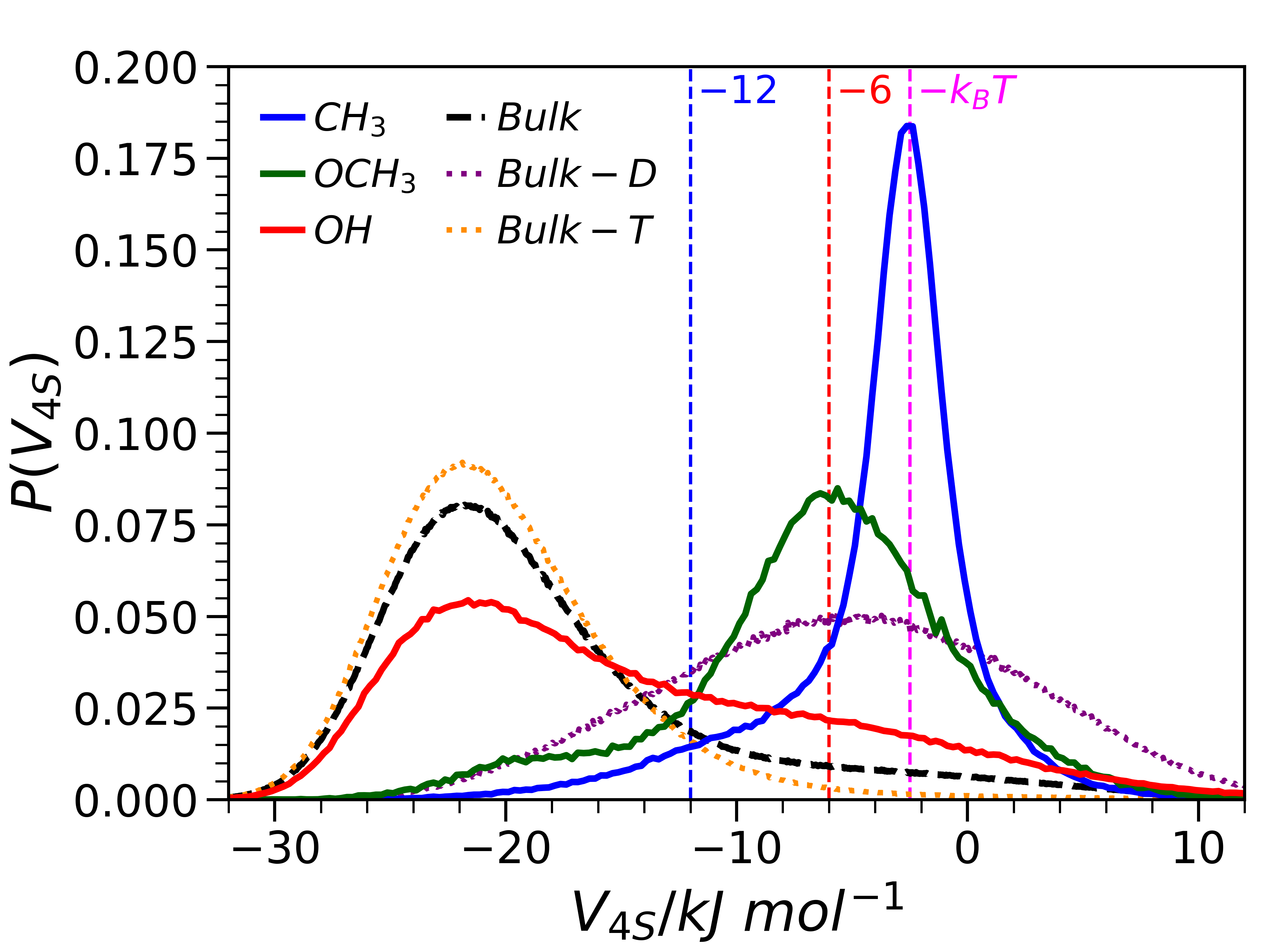}
}
\captionsetup{font=small} 
\caption{$V_{4S}$ distributions for three SAMs (ending in OH, $\mathrm{CH_3}$ and $\mathrm{OCH_3}$). We also include the distribution for bulk water (the population of T molecules is dominat, thus the prevalence of the left-side peak for the whole bulk distribution depicted by the dashed black line) and its decomposition in T and D molecules (normalized to  unitary values). We use TIP4P/2005 and temperature T=300K.}
\label{fig1}
\end{figure}

Recently, we have shown that water molecules present two interaction requirements that govern the structuring of the liquid state: on the one hand, the establishment of linear hydrogen bonds that demand the well-known local expansion of its second molecular shell and, on the other hand, the local contraction of the second shell to partially compensate for HB-lacking sites\cite{v4s1,v4s2,arxiv}. The former is responsible for the establishment of liquid water's extended HB network (improving energy by structure expansion, in contrast with the situation in normal liquids). In turn, the latter allows for the existence of HB coordination defects, which would be otherwise exceedingly costly within the network's expanded environment. Hence, this structure-energy fine-tuning materializes itself into two competing local molecular arrangements: tetrahedral (four-fold HB-coordinated) molecules, which we named as T molecules, and defective (HB-undercoordinated) molecules, termed as D ones. The radial distribution functions of these two kinds of molecules have been shown to closely resemble that of Low-Density Amorphous ice (LDA) and High-Density Amorphous ice (HDA), respectively\cite{v4,v4T2}. We note, however, that while our present approach is consistent with the liquid-liquid critical point (LLCP) scenario\cite{water_chemrev,water1_epje,water2_epje,criticalpoint,Kim,francesco-pablo} (that extends the HDA-LDA coexistence line to the liquid regime), it only relies on the existence of the D and T molecules (HDA-like and LDA-like molecules, respectively). It is worth mentioning that it has recently been asked whether the LLCP, which has been extremely successful in describing water anomalies, would help us describe water's behavior beyond the supercooled regime, precisely aiming at its ``interactions" in ``real-life” situations\cite{FinneyLLCP}. Within this context, an energetically-based water structure indicator, valid both for bulk and non-bulk contexts, was built\cite{v4s1,v4s2,arxiv}. This index was called $V_{4S}$ given that it was inspired by the directional nature of water interactions. Basically\cite{v4s1,v4s2,arxiv}, it considers four tetrahedrally-arranged interaction sites for the water molecule, adds up all the different interactions felt at each of these sites and orders them from the strongest interacting ($V_{1S}$) to the weakest one ($V_{4S}$) (see the Supplementary Material and https://github.com/nicolas-loubet/V4S for details). For bulk water (as shown in Fig.~\ref{fig1}), $V_{4S}$ displays a bimodal distribution resulting form a left peak at a value consistent with the energy of a linear HB and a broader right peak at around -6kJ/mol. $V_{1S}$, $V_{2S}$ and $V_{3S}$ only display a peak within the linear HB region. The 4-fold HB coordinated T molecules were found to contribute to the left peak of the $V_{4S}$ distribution, while the (minority) 3-fold HB coordinated D molecules were responsible for the (rather temperature-independent) right peak at -6kJ/mol\cite{v4s1,v4s2,arxiv} (see Fig.~\ref{fig1}). The peak at the left is almost completely dominated by a linear HB (first molecular shell) since the second molecular shell has expanded to allow for HB improvement and does not interact significantly with the central molecule. The peak on the right, in turn, results from attractions with usually more than one molecule of the second shell, which has locally contracted to partially compensate the otherwise costly lacking-HB, a compensation that is quite significant as compared to $k_B T$ (see\cite{v4s1,v4s2} and, more specifically \cite{arxiv}). This clearly two-state nature of water interactions has been shown, thus, to be central for its peculiar structuring. Beyond bulk environments, application of $V_{4S}$ to water hydrating certain typical hydrophilic and hydrophobic surfaces enabled to discriminate their bahavior, since water molecules in contact with the former presented low index values, while molecules close to the latter displayed values larger than that of bulk D molecules\cite{v4s1,v4s2,arxiv}. 

On the basis of the above-expounded observations, at this point we pose the following tenet: Given both the reluctance of water to lose HB-coordination and its demand for partial compensation of HB-lacking sites, the molecular principle that rules bulk-like structuring would also hold valid in order to establish favorable interactions with other systems. If this were true, these two interaction requirements (that enable the local growth of water's HB network together with its defects) would translate into two limiting wetting behaviors: A hydrophilicity saturation limit and a hydrophilicity threshold marking the transition to the hydrophobic regime. We shall hereby show that this is indeed the case for surfaces functionalized to render a variable degree of hydrophilicity. In this sense, our results will provide strong support for the existence of the hitherto elusive molecular signature of hydrophilicity expected to set in below a $\theta=90 ^{\circ}$ contact angle value, while also enabling the establishment of an absolute hydrophobicity/hydrophilicity scale.

To test our proposition, we shall use a battery of self-assembled monolayers (arrangements of alkane-like chains), whose degree of hydrophilicity can be varied from the hydrophilic to the hydrophobic regime by proper functionalization of their end-chains in contact with water, and for which both experimental\cite{SAMS_exp} and theoretical\cite{Garde_PNAS} contact angle calculations have been performed in the past (details of the simulations can be found in the Supplementary Material, and in previous work\cite{v4s1,v4s2,arxiv}). Indeed, this represents an excellent test system since a rather large spectrum of hydrophobicity/hydrophilicity is covered and, more conveniently, one of the surfaces (the metoxi-functionalized SAM, $\mathrm{SAM-O CH_3}$) presents an experimental contact angle just below $90 ^{\circ}$ (that is, very slightly hydrophilic) close to the expected hydrophilicity threshold the water structure index would predict. Fig.~\ref{fig1} displays the distributions of the $V_{4S}$ index for the water molecules close to three of the SAMs studied: methyl-, hydroxyl- and metoxi-functionalized SAMS. In all cases we consider water molecules closer than 4\AA\ to any heavy atom of the end groups of the SAM chains (we discard some water molecules within such region not in direct contact with the surface, that whose all $V_{1S}$ to $V_{4S}$ are given by water-water interactions lower than -12kJ/mol, consistent with a HB). As already indicated, we also show the $V_{4S}$ distribution for bulk water (dashed line) and the cases for the separated contributions to bulk behavior of the D and T molecules when the populations of both classes are normalized to unitary value. As expected, the index performs well in discriminating the hydrophilic SAM, $\mathrm{SAM-OH}$, from the hydrophobic one, $\mathrm{SAM-CH_3}$. While the former presents a distribution close to that of bulk water (implying the formation of good SAM-water HBs), the latter presents a peak located to the right as compared to both kinds of bulk water molecules, slightly above a value corresponding to a $k_B T$ attraction. The very slightly hydrophilic $\mathrm{SAM-O CH_3}$, in turn, presents a distribution similar to that of the bulk D molecules, with a mean value slightly above their corresponding one (-6kJ/mol). This represents a good first hint towards the validity of our approach on a quantitative basis.

\begin{figure}[h!]
\resizebox{0.5\textwidth}{!}{%
\includegraphics{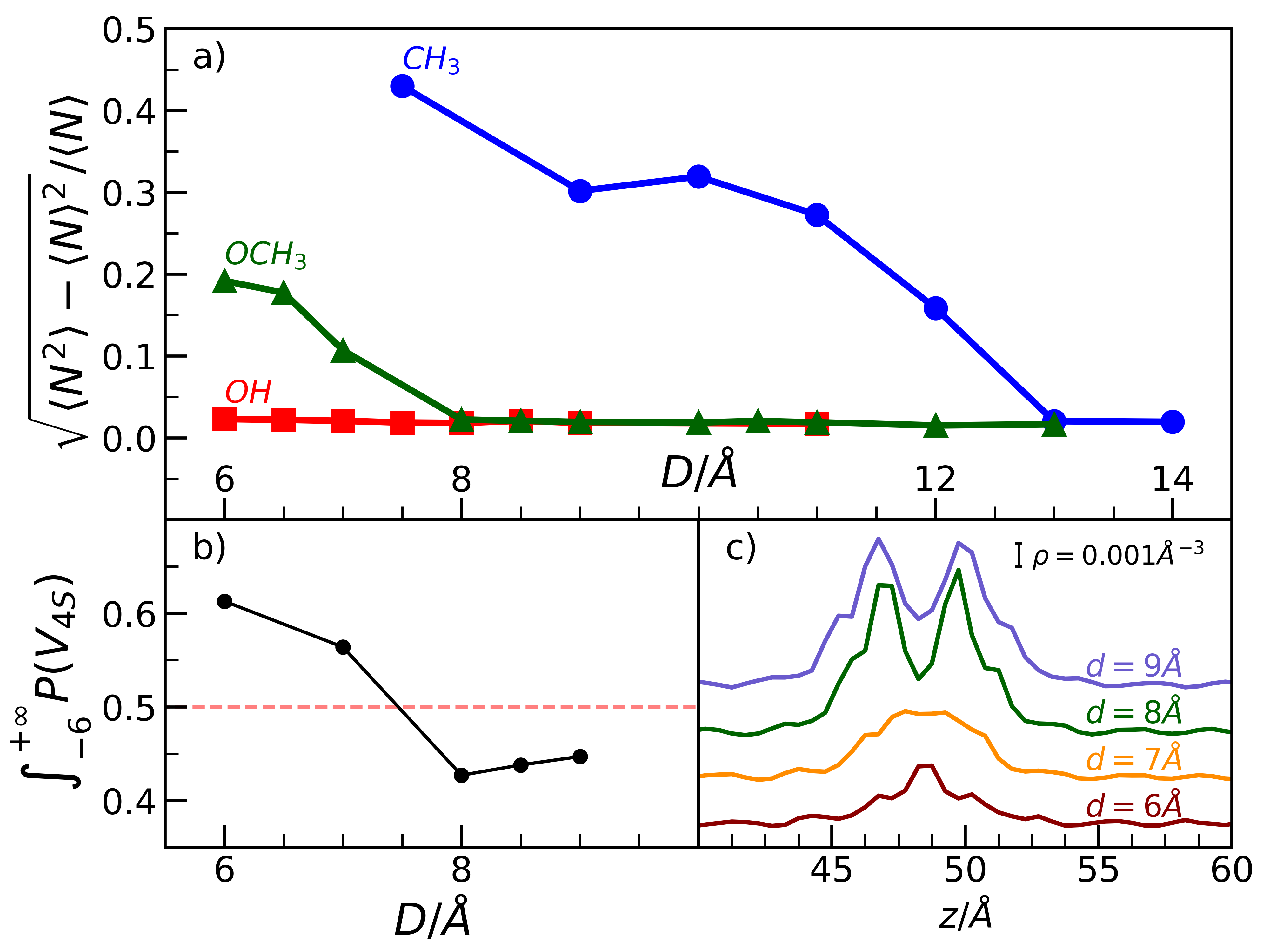}
}
\captionsetup{font=small} 
\caption{a) Water number density fluctuations (ratio of the dispersion to the average value) as a function of plate separation for three SAMs; b) fraction of  water molecules with $V_{4S}$ corresponding to hydrophobic behavior (higher than -6.0kJ/mol) as a function of plate separation for the $\mathrm{SAM-OCH_3}$; c) Water density profile normal to the plates within the interplate region of the $\mathrm{SAM-OCH_3}$ at different indicated separations.}
\label{fig2}
\end{figure}
 
To better quantify the performance of the index, we also studied systems of two parallel plates built by the same SAM systems already considered (pairs of identically functionalized SAMs exposing their functional groups to the interplate water region), at different separations. 
In Fig.~\ref{fig2} we show the behavior of the index in the interplate region together with that of the normalized water number density fluctuations within such zone. Water density fluctuations are expected to be large close to a hydrophobic surface since such water molecules would be easily removable from the observation volume. Hydrophilic surfaces, in turn, should display small values since the water molecules are tightly bound to the surface. As expected, for large plates separation the water density fluctuations for the three SAM systems ($\mathrm{SAM-CH_3}$, $\mathrm{SAM-OH}$ and $\mathrm{SAM- OCH_3}$) display a low value consistent with bulk-like behavior. This behavior is roughly independent of plate separation for the hydrophilic $\mathrm{SAM-OH}$, since the water molecules are tightly bound to the plate surfaces. For the hydrophobic $\mathrm{SAM-CH_3}$, the fluctuations begin to significantly increase as separation decreases, since bulk-like water (water molecules surrounded only by other water molecules) starts to no longer be present, as all confined molecules get close to the hydrophobic plates and sacrifice hydrogen bonding. In turn, in the case of the $\mathrm{SAM-OCH_3}$, a hydrophobic regime only sets when there is a transition from two to one water layers between the plates (as evident from the plot of the interplate water density profiles normal to the $\mathrm{SAM-OCH_3}$ surfaces). Hhydrophilic behavior is lost below this separation, since (unlike for the $\mathrm{SAM-OH}$) the slightly hydrophilic nature of the surface is on the verge of the hydrophilic/hydrophobic limit and the water-wall interactions are perturbed by the extreme confinement. To make this fact more explicit, we also include a calculation of the fraction of water molecules with $V_{4S}$ consistent with hydrophobic values ($V_{4S}$ higher than -6.0kJ/mol) for different separations of the $\mathrm{SAM-OCH_3}$ plates.  When the single monolayer regime is reached (below around 8\AA), the fraction of high index molecules crosses 0.5 and, more notably, this fraction steadily grows as separation decreases, since the extreme confinement begins to spoil the interaction geometries in a system for which the interaction strength was already very slightly hydrophilic.

\begin{figure}[h!]
\resizebox{0.5\textwidth}{!}{%
\includegraphics{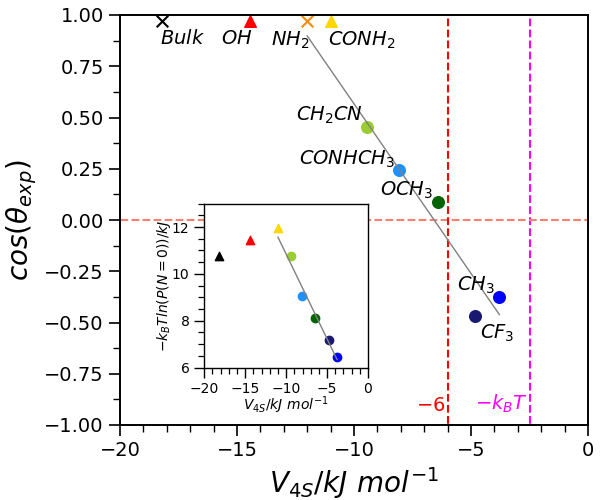}
}
\captionsetup{font=small} 
\caption{Correlation of the experimentally-measured contact angle ($\theta$, extracted from\cite{SAMS_exp,Garde_PNAS}) with $V_{4S}$ for a set of SAMs (ended in $\mathrm{CH_3}$, $\mathrm{CF_3}$, $\mathrm{OCH_3}$, $\mathrm{CONHCH_3}$, $\mathrm{CH_2CN}$, $\mathrm{CONH_2}$ and $\mathrm{OH}$). We also include a $\mathrm{SAM-NH_2}$ (for which there is no experimental value available but that should experience strong hydrophilic behavior) and the corresponding behavior for bulk water. Inset: Correlation of the water vacating probability (P(N=0), simulation values extracted from\cite{Garde_PNAS}) with $V_{4S}$. In both cases, extrapolation lines are for systems outside the strong hydrophilicity saturation limit (circles). Crosses ($\mathrm{SAM-NH_2}$ and bulk water) indicate values only from simulations.}
\label{fig3}
\end{figure}

Having shown that the $V_{4S}$ index provides an appropriate measure of hydrophilicity, in Fig.~\ref{fig3} we correlate the experimental contact angles (extracted from\cite{SAMS_exp,Garde_PNAS}) with the mean $V_{4S}$ values for a set of different SAMs. We also consider an amine functionalized SAM ($\mathrm{SAM-NH_2}$, for which we do not have the experimental value of its contact angle, $\theta$) and include the mean $V_{4S}$ value of bulk water. All the experimentally hydrophobic and hydrophilic SAMs are correctly classified by $V_{4S}$. Moreover, a nice linear correlation is obtained. For the hydrophobic SAMs, the value of the index is higher than -6kJ/mol (the value for bulk-like defective D molecules) a threshold consistent with the zero value for $\cos (\theta)$ (that is, $\theta=90^{\circ}$). As this value is crossed towards lower $V_{4S}$ values (the wall-water interactions are stronger), the systems belong to the hydrophilic regime. This behavior saturates when the limit of strong hydrophilicity is reached at $\theta=0^{\circ}$. This limit occurs when a good linear HB is formed, consistent with an interaction $V_{4S} \leq -12 kJ/mol$ which implies an environment consistent with that of a T molecule. Thus, this plot shows that not only the structural index successfully classifies all the SAMs as hydrophobic or hydrophilic as in the contact angle experiments, it also displays a quantitative hydrophobicity/hydrophilicity scale consistent with experiments and, most importantly, enables one to identify the molecular signature of the transition that occurs at $\theta=90^{\circ}$: Contact angles below such value imply that the surface is able to provide the local hydrating water molecules with an environment at least energetically equivalent to that of the bulk D molecules, while values above $\theta=90^{\circ}$ are a result of interactions that are not strong enough to compete with a bulk-like environment. As an inset, we also correlate the $V_{4S}$ values with the water vacating probability, P(N=0), extracted from the work of Garde's group\cite{Garde_PNAS}. The probability P(N) of finding N water molecules inside a sphere of radius r=3.3\AA\ tangent to the surface is related to the fluctuations in number density. Hydrophobic behavior is indicated by large values of P(N = 0), which is inversely proportional to the work necessary to completely remove water molecules from the sphere, since $-k_B T \ln P(N = 0)$ can be understood as the energy spent to create such a cavity). P(N=0) represents thus a practical operational hydrophobicity measure even when it cannot detect a transition between hydrophobic and hydrophilic behavior. A remarkable correlation with $V_{4S}$ is also evident from such plot.

In this work we have shown that the two interaction demands that rule bulk-water structuring are also relevant for rationalizing wetting behavior. In this sense, if an interacting extensive surface is able to fully replace water-water HBs, the system reaches a strong hydrophilic limit characterized by a contact angle $\theta \approx 0^{\circ}$. When the surface cannot avoid disruption of water's HB network, partial hydrophilic behavior still holds if the lacking HBs are compensated at least at the level they attain in bulk-like conditions. Finally, when the $V_{4S}$ structural indicator shows that the water-wall interactions start to stop being favorable, the contact angle crosses $\theta = 90^{\circ}$ towards larger values typical of the hydrophobic regime. The water molecules still hydrate such an extensive hydrophobic surface, but suffer fast interchanges with the bulk where they can satisfy the interaction demands unfulfilled when close to the surface (thus enhancing the translational diffusivity)\cite{v4s2}. Thus, a careful consideration of the subtleties involved in water structuring and interaction has enabled not only to reveal the molecular underpinnings of hydrophilicity, but allowed to endow this notion with absolute values by relating a widely used macroscopic/mesoscopic measure, the contact angle, to a novel molecular-scale indicator.

\end{document}